\documentstyle[12pt,psfig,aasms4]{article}

\newcommand{\CVT}{ {\scriptscriptstyle {CVT}} }

\newcommand{\gtsima}{$\; \buildrel > \over \sim \;$}
\newcommand{\ltsima}{$\; \buildrel < \over \sim \;$}
\newcommand{\simgt}{\lower.5ex\hbox{\gtsima}}
\newcommand{\simlt}{\lower.5ex\hbox{\ltsima}}
\newcommand{\hikpc}{{\hbox {$h^{-1}$}{\rm kpc}} }
\newcommand{\himpc}{{\hbox {$h^{-1}$}{\rm Mpc}} }
\newcommand{\kms}{ {\rm km/sec} }

\newcommand{\0}{ {\scriptscriptstyle 0} }

\def\pp{\par\parshape 2 0truecm 15.5truecm 1truecm 14.5truecm\noindent}
\begin{document}
\baselineskip=20pt

\vspace*{-1.5cm}
\begin{minipage}[c]{3cm}
  \psfig{figure=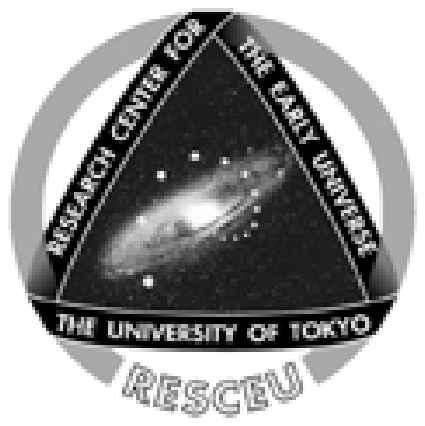,height=3cm}
\end{minipage}
\begin{minipage}[c]{9cm}
\begin{centering}
{
\vskip 0.1in
{\large \sf 
THE UNIVERSITY OF TOKYO\\
\vskip 0.1in
Research Center for the Early Universe}\\
}
\end{centering}
\end{minipage}
\begin{minipage}[c]{3cm}
\vspace{2.5cm}
RESCEU-33/96\\
UTAP-240/96
\end{minipage}\\
\vspace{1cm}

\title{The finite size effect of galaxies\\
 on the cosmic virial theorem\\
and the pairwise peculiar velocity dispersions}

\bigskip

\author{Yasushi Suto$^{1,2}$ ~and~ Yi-Peng Jing$^{1,3}$}

\bigskip
\bigskip

\affil{
$^{1}$ Max-Planck-Institut f\"{u}r Astrophysik, 
Karl-Schwarzschild-Strasse 1, \\ 85748 Garching, Germany\\
$^{2}$ permanent address: Department of Physics and RESCEU \\
(Research Center for the Early Universe), 
School of Science, \\ The University of Tokyo, Tokyo 113, Japan\\
$^{3}$ address after November 1, 1996. RESCEU (Research Center for the
Early Universe),\\ School of Science, The University of Tokyo, Tokyo
113, Japan\\
e-mail: suto@phys.s.u-tokyo.ac.jp, 
jing@utaphp2.phys.s.u-tokyo.ac.jp\\ }

\received{1996 August 22}

\baselineskip=20pt

\begin{abstract}
  We discuss the effect of the finite size of galaxies on estimating
  small-scale relative pairwise peculiar velocity dispersions from the
  cosmic virial theorem (CVT). Specifically we evaluate the effect by
  incorporating the finite core radius $r_c$ in the two-point
  correlation function of mass, i.e. $\xi_\rho(r) \propto
  (r+r_c)^{-\gamma}$ and the effective gravitational force softening
  $r_s$ on small scales.  We analytically obtain the lowest-order
  correction term for $\gamma <2$ which is in quantitative agreement
  with the full numerical evaluation. With a nonzero $r_s$ and/or
  $r_c$ the cosmic virial theorem is no longer limited to the case of
  $\gamma<2$. We present accurate fitting formulae for the CVT
  predicted pairwise velocity dispersion for the case of
  $\gamma>2$. Compared with the idealistic point-mass approximation
  ($r_s=r_c=0$), the finite size effect can significantly reduce the
  small-scale velocity dispersions of galaxies at scales much larger
  than $r_s$ and $r_c$.  Even without considering the finite size of
  galaxies, nonzero values for $r_c$ are generally expected, for
  instance, for cold dark matter (CDM) models with a scale-invariant
  primordial spectrum. For these CDM models, a reasonable force
  softening $r_s\le 100 \hikpc$ would have rather tiny effect. We
  present the CVT predictions for the small-scale pairwise velocity
  dispersion in the CDM models normalized by the COBE observation.
  The implication of our results for confrontation of observations of
  galaxy pair-wise velocity dispersions and theoretical predictions of
  the CVT is also discussed.
\end{abstract}


\baselineskip=20pt

\section{INTRODUCTION}

The three-dimensional distribution of galaxies observed in redshift
surveys differs from the true one since the distance to each galaxy
cannot be determined by its redshift $z$ only; for $z \ll 1$ the
peculiar velocity of galaxies, typically $\sim (100-1000)\kms$,
contaminates the true recession velocity of the Hubble flow (e.g.,
Davis \& Peebles 1983; Kaiser 1987; Hamilton 1992), while the true
distance for objects at $z\simgt1$ sensitively depends on the (unknown
and thus assumed) cosmological parameters. This hampers the effort to
understand the true distribution of large-scale structure of the
universe. Nevertheless such redshift-space distortion effects are
quite useful since through the detailed theoretical modeling, one can
derive the peculiar velocity dispersions of galaxies as a function of
separation (Davis \& Peebles 1983), and also can infer the
cosmological density parameter $\Omega_0$ and the dimensionless
cosmological constant $\lambda_0$, for instance (Ballinger, Peacock \&
Heavens 1996; Matsubara \& Suto 1996; Suto \& Matsubara 1996).

Theoretically, a conventional tool to predict small-scale relative
pairwise peculiar velocity dispersions is the cosmic virial theorem
(Peebles 1976; hereafter CVT). In its simplest form, the prediction is
based on the idealistic assumption that galaxies are treated as point
particles. As Peebles (1976) remarked, however, the finite size effect
is very significant on scale $\sim 1\himpc$ even if one considers a
radius of $(20\sim100) \hikpc$ for typical galactic halos. In this
paper we explore this realistic effect in more details and discuss its
important impact on the application of the CVT to real
observations. We will also give reliable predictions for the
small-scale velocity dispersions of galaxies in COBE normalized CDM
models.

The plan of the paper is as follows; section 2 briefly describes the
CVT both in its simplest and in its more realistic form. The finite size
effect is considered in \S 3, first based on a perturbation theory (\S
3.1) and then on a full numerical analysis (\S 3.2).
In \S 4 we predict velocity dispersion of galaxies
on small scales for COBE normalized CDM models. Discussion and
implications of our result, and our main conclusions are given 
in \S 5.

\section{COSMIC VIRIAL THEOREM}

Assuming that galaxy pairs are gravitationally in ``statistical
equilibrium'' on small scales, Peebles (1976, 1980) derived an
expression for their relative (one-dimensional) peculiar velocity
dispersion as a function of their separation $r$:
\begin{equation}
\langle v^2_{12} (r) \rangle_\CVT
  = {6 G \bar \rho \over \xi_\rho(r)} 
   \int_r^\infty {d r' \over r'} 
   \int d {\bf z} \, { {\bf  r}' \cdot {\bf z} \over z^3} 
    \zeta_\rho (r', z, |{\bf  r}' - {\bf  z}|) ,
\label{eq:cvt1}
\end{equation}
where $\bar \rho$ is the mean density of the universe, and
$\xi_{\rho}(r)$ and $\zeta_{\rho}$ are the two- and three-point
correlation functions of {\it mass}.

The observed two- and three-point correlation functions of {\it
  galaxies}, $\xi_{\rm g}$ and $\zeta_{\rm g}$, are well approximated
by the following forms (Groth \& Peebles 1977; Davis \& Peebles 1983):
\begin{eqnarray}
  \xi_{\rm g}(r) &=& \left( {r_\0 \over r}\right)^\gamma , \\ 
  \zeta_{\rm g} (r_1, r_2, r_3) &=& Q_{\rm g} \, [\xi_{\rm
    g}(r_1)\xi_{\rm g}(r_2) + \xi_{\rm g}(r_2)\xi_{\rm g}(r_3) +
  \xi_{\rm g}(r_3)\xi_{\rm g}(r_1)]
\end{eqnarray}
where $r_0=(5.4\pm0.3)\himpc$, $\gamma=1.77\pm0.04$, $Q_{\rm
g}=1.29\pm 0.21$ ($h$ is the dimensionless Hubble constant $H_0$ in
units of $100\;{\rm km \, s^{-1}\, Mpc^{-1}}$).  Thus it is reasonable
to assume that the two- and three-point correlation functions of {\it
mass}, $\xi_\rho$ and $\zeta_\rho$, also obey the same scaling except
for the overall amplitudes:
\begin{eqnarray}
  \xi_\rho(r) = {1 \over b_{\rm g}^2} \xi_{\rm g}(r), \qquad
  \zeta_\rho (r_1, r_2, r_3) = {Q_\rho \over Q_{\rm g} b_{\rm g}^4 }
  \zeta_{\rm g} (r_1, r_2, r_3) .
\label{eq:hier}
\end{eqnarray}
Then a straightforward computation yields
\begin{equation}
\langle v^2_{12} (r) \rangle_\CVT 
  = 6 G \bar \rho Q_\rho \, r_\0^\gamma \, r^{2-\gamma} \, I_0(\gamma)/b_g^2,
\end{equation}
where
\begin{eqnarray}
&&I_0(\gamma) \equiv {\pi \over (\gamma-1)(2-\gamma)(4-\gamma)} \nonumber \\
&&\times \int_0^\infty \, dx\, {1+x^{-\gamma} \over x^2} 
\left\{ 
(1+x)^{4-\gamma}-|1+x|^{4-\gamma}
-(4-\gamma)x\left[(1+x)^{2-\gamma}+|1+x|^{2-\gamma}\right]
\right\} .
\label{eq:i0gamma}
\end{eqnarray}
Numerically $I_0(1.65)\sim 25.4$, $I_0(1.8)\sim 33.2$, and
$I_0(1.95)\sim 55.6$. More specifically the CVT predicts the
small-scale peculiar velocity dispersion as (Peebles 1976; Suto 1993)
\begin{eqnarray}
\langle v^2_{12} (r) \rangle^{1/2}_\CVT = 1460
     \sqrt{\Omega_0 Q_\rho \over 1.3 b_{\rm g}^2}
  \sqrt{I_0(\gamma) \over 33.2} 5.4^{\frac {\gamma-1.8}2} \left({r_\0
      \over 5.4\himpc} \right)^{\frac\gamma 2} \left({r \over 1\himpc }
  \right)^{\frac{2-\gamma}2} \kms .
\label{eq:cvt1d}
\end{eqnarray}

The above simple estimate can be improved more realistically by
including the finite size of the galaxies, which suppresses the
effective gravitational force between pairs, and also possibly changes
the behavior of the two-point correlation functions on small scales.
In fact Peebles (1976) already remarked that the above two effects may
change the predictions of the velocity dispersions significantly. More
recently Bartlett \& Blanchard (1996) also discussed the importance of
the local mass distribution around pairs in the CVT.  Our principal
aim here is to describe these effects more extensively using both the
perturbation analysis and the full numerical integration. For this
purpose we take account of the former effect by softening the
gravitational force according to the Plummer law with the softening
radius $r_s$. Then equation (\ref{eq:cvt1}) is modified as
\begin{equation}
\langle v^2_{12} (r) \rangle_\CVT
  = {6 G \bar \rho \over \xi_\rho(r)} 
   \int_r^\infty {d r' \over r'} 
   \int d {\bf  z} \, 
{ {\bf r}' \cdot {\bf z} \over (z^2+r_s^2)^{3/2}} 
    \zeta_\rho (r', z, |{\bf  r}' - {\bf  z}|) .
 \label{eq:cvtrs}
\end{equation}
We model the latter effect by incorporating a nonzero core radius
$r_c$ as follows:
\begin{eqnarray}
  \xi_{\rm g}(r) = \left( {r_\0 \over r+r_c}\right)^\gamma .
\label{eq:xirc}
\end{eqnarray}
The two modifications can be still described in equation
(\ref{eq:cvt1d}) simply by replacing $I_0(\gamma)$ with
\begin{eqnarray}
&&I(y, w ; \gamma) \equiv (1+y)^\gamma 
\int {d {\bf t} \over (t^2+w^2)^{3/2}} 
 \int_1^\infty {d s \over s} \, 
 { {\bf  s} \cdot {\bf  t }
  \over (|{\bf  s} - {\bf  t}| + y)^\gamma} 
 \left[ {1 \over (s+y)^\gamma} + {1 \over (t+y)^\gamma}\right]
\nonumber \\
&&= 2\pi (1+y)^\gamma 
\int_0^\infty {t^3 dt \over (t^2+w^2)^{3/2}} 
 \int_1^\infty ds
 \left[ {1 \over (s+y)^\gamma} + {1 \over (t+y)^\gamma}\right] 
K(s,t,y;\gamma) ,
\label{eq:iyw}
\end{eqnarray}
where $y\equiv r_c/r$, $w\equiv r_s/r$, and
\begin{eqnarray}
K(s,t,y;\gamma) &\equiv& \int_{-1}^{1} {\mu d\mu \over
(\sqrt{s^2+t^2-2st\mu} + y)^\gamma} \nonumber \\
&=& 
{(s+t+y)^{4-\gamma} - (|s-t|+y)^{4-\gamma} \over 2s^2t^2 (\gamma-4)}
 \nonumber \\
&+& {3y \over 2s^2t^2 (3-\gamma)} 
\left[(s+t+y)^{3-\gamma} - (|s-t|+y)^{3-\gamma}\right] \nonumber \\
&+& {s^2+t^2-3y^3 \over 2s^2t^2 (2-\gamma)} 
\left[(s+t+y)^{2-\gamma} - (|s-t|+y)^{2-\gamma}\right]
 \nonumber \\
&+& {y(y^2-s^2-t^2) \over 2s^2t^2 (1-\gamma)} 
\left[(s+t+y)^{1-\gamma} - (|s-t|+y)^{1-\gamma}\right] .
\end{eqnarray}

The integral $I(y,w;\gamma)$ is divergent for $\gamma>2$ if $y=w=0$.
Nonzero values for $y$ and/or $w$ make $I(y,w;\gamma)$ finite for
$\gamma >2$, therefore incorporating the finite size of galaxies in
this way widens the applicability of the CVT compared with its
simplest form (\ref{eq:cvt1d}). For realistic CDM models, the
mass two-point correlation functions on small scales may be better 
described by equation
(\ref{eq:xirc}) with $\gamma \ga 2$ and $r_c>0$, as will be seen
below.

\section{EFFECT OF THE SIZE OF GALAXIES}

\subsection{perturbation analysis for $\gamma<2$}

In the case of $\gamma<2$, the integral $I(y,w;\gamma)$ converges even
for $y=w=0$. Then one may treat the effect of nonzero $y$ and $w$ via
perturbation theory. Consider the case of $w=0$ first. Let us set $y =
u^\alpha$ ($\alpha$ is a parameter greater than unity and will be
determined below), and try to find the Taylor expansion with respect
to $u$:
\begin{eqnarray}
I(u^\alpha,0;\gamma) = I(0,0;\gamma) 
+ \left.{\partial I \over \partial u}\right|_{u=0}u  + O(u^2) .
\end{eqnarray}
The derivative $\partial I/\partial u$ is explicitly given as
\begin{eqnarray}
&&{1 \over 2\pi} {\partial I \over \partial u}
= \alpha\gamma u^{\alpha-1}(1+u^\alpha)^{\gamma-1}
\int_0^\infty dt \int_1^\infty ds
 \left[ {1 \over (s+u^\alpha)^\gamma} + 
{1 \over (t+u^\alpha)^\gamma}\right] K(s,t,u^\alpha;\gamma) 
\nonumber\\
&-& \alpha\gamma u^{\alpha-1} (1+u^\alpha)^\gamma
\int_0^\infty dt \int_1^\infty ds
 \left[ {1 \over (s+u^\alpha)^{\gamma+1}} + 
{1 \over (t+u^\alpha)^{\gamma+1}}\right] K(s,t,u^\alpha;\gamma) 
\nonumber \\
&-& \alpha\gamma u^{\alpha-1} (1+u^\alpha)^\gamma
\int_0^\infty dt \int_1^\infty ds
 \left[ {1 \over (s+u^\alpha)^{\gamma}} + 
{1 \over (t+u^\alpha)^{\gamma}}\right] K(s,t,u^\alpha;\gamma+1) .
\end{eqnarray}
In the limit of $u \rightarrow 0$, only the second term gives a
non-vanishing contribution:
\begin{eqnarray}
{1 \over 2\pi}\left. {\partial I \over \partial u}\right|_{u=0}
= - \lim_{u\rightarrow 0}
 \alpha\gamma u^{\alpha-1} 
\int_0^\infty dt \int_1^\infty ds
{1 \over (t+u^\alpha)^{\gamma+1}}
\left[ {2\gamma \over 3(s+u^\alpha)^{\gamma+1}} t + O(t^2) \right] .
\label{eq:iulim}
\end{eqnarray}
By setting $t\equiv u^\alpha\tau$ and then considering the limit $u
\rightarrow 0$, one finds that only the choice of $\alpha =
1/(2-\gamma)$ gives the finite value for equation (\ref{eq:iulim})
which is contributed only by the term proportional to $t$ in the square
bracket:
\begin{eqnarray}
{1 \over 2\pi}\left.{\partial I \over \partial u}\right|_{u=0}
= - \alpha\gamma
\int_1^\infty 
{2\gamma ds \over 3 s^{\gamma+1}}
\int_0^\infty {\tau  d\tau \over (\tau +1)^{\gamma+1}}
= - {2 \over 3(\gamma-1)(2-\gamma)}
\qquad (1<\gamma<2)
\end{eqnarray}

The above method can be also applied to the perturbative analysis for
$I(0,w;\gamma)$ and $I(y,w=y;\gamma)$, and a straightforward
calculation yields
\begin{eqnarray}
I(y,0;\gamma) &=& I(0,0;\gamma) - {4\pi \over 3(\gamma-1)(2-\gamma)}
\, y^{2-\gamma} + O(y^{4-2\gamma}), \label{eq:iy0} \\
I(0,w;\gamma) &=& I(0,0;\gamma) - {2\pi \over 2-\gamma}
B({5-\gamma \over 2}, {\gamma \over 2})
\,  w^{2-\gamma} + O(w^{4-2\gamma}) , \label{eq:i0w} \\
I(y,y;\gamma) &=& I(0,0;\gamma) 
- {4\pi \,  y^{2-\gamma} \over 3(2-\gamma)}
\int_0^\infty {\tau^4 (\gamma \tau^2 +3\tau+\gamma+3) \over (\tau^2+1)^{5/2}
(\tau+1)^{\gamma+1} } d\tau + O(y^{4-2\gamma}) ,
\label{eq:iyy}
\end{eqnarray}
where $B(\alpha,\beta)$ is the beta function. The above expressions
clearly show the importance of the finite size effect for the
observational value $\gamma=1.8$. Since the lowest-order correction
term begins with $y^{2-\gamma}$, even a very small value of $y$
changes the value of $I(y,w;\gamma)$ significantly, which will be
discussed in details below.

\subsection{numerical evaluation}

\begin{figure}
\vspace*{-1.0cm}
\begin{center}
   \leavevmode\psfig{figure=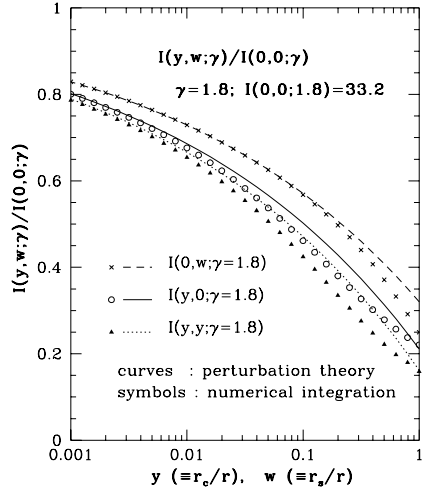,width=7.5cm}
\end{center}
\vspace*{-1.0cm}
\caption{Effect of nonzero $y$ and $w$ on $I(y,w;\gamma)$ for 
$\gamma=1.8$. Symbols indicate the results of direct numerical 
evaluations while thin curves plot those of the perturbation theory.
\label{fig:iywg18}
}\end{figure}

Figure \ref{fig:iywg18} plots the results of the numerical evaluation
of equation (\ref{eq:iyw}) for $\gamma=1.8$ in comparison with the
perturbative expansions (\ref{eq:iy0}) to (\ref{eq:iyy}).  In
practice, here and in what follows we set $w=10^{-10}$ and
$y=10^{-10}$, respectively, when we numerically compute the values of
$I(y,0;\gamma)$ and $I(0,w;\gamma)$.  The perturbation result is in
excellent agreement with the numerical evaluation up to $y=0.001$ for
$I(y,0;1.8)$ and $I(y,y;1.8)$, and up to $w=0.1$ for $I(0,w;1.8)$.
Also the perturbation correction turns out to provide a reasonably
good approximation even up to $y\sim w\sim 1$.  Thus for
astrophysically relevant values of $r_s$ and $r_c$, the perturbation
formulae (eqs.[\ref{eq:iy0}] to [\ref{eq:iyy}]) are very useful if $1<
\gamma < 2$.  It should be noted that even very small values of $y$
and/or $w$ significantly decrease $I(y,w,\gamma)$ relative to the
idealistic point-mass approximation $I(0,0,\gamma)=I_0(\gamma)$; for
$\gamma=1.8$, $I(y,w,\gamma)/I(0,0,\gamma)$ is about 0.8 even if $y$
or $w$ is $\sim 10^{-3}$. On the other hand, the dependence on $y$ and
$w$ is fairly weak as indicated by the perturbation correction
(\ref{eq:iy0}) to (\ref{eq:iyy}). Therefore in a sense most of the
effect might be interpreted as the inherent offset originated from the
point-mass approximation itself.

\begin{figure}
\vspace*{-1.0cm}
\begin{center}
   \leavevmode\psfig{figure=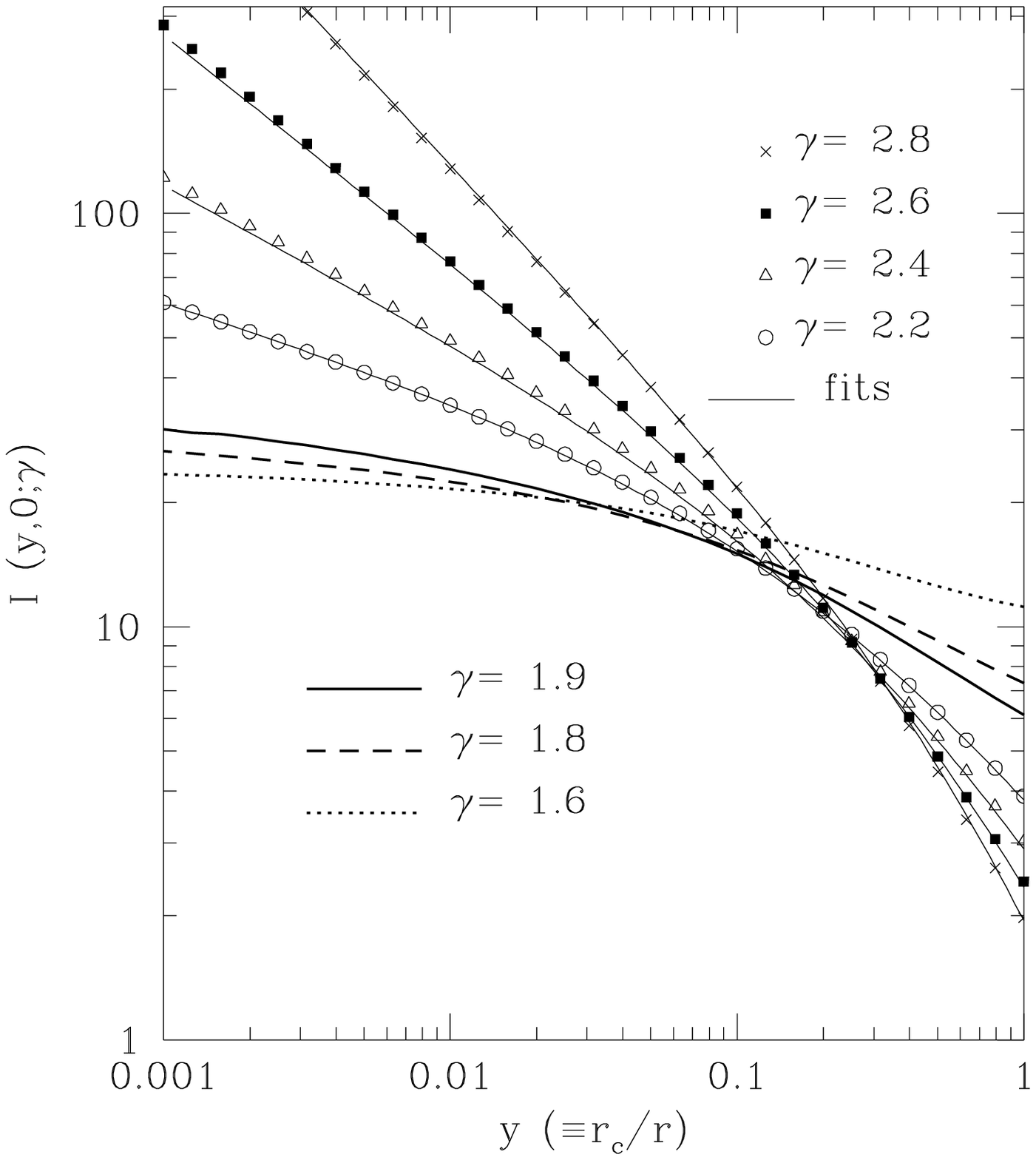,width=7.5cm}
\end{center}
\vspace*{-1.0cm}
\caption{ $I(y,0;\gamma)$ for different $\gamma$ ($r_s=0$).  Thick
curves indicate for $\gamma<2$ while symbols (direct numerical
integration) and thin solid curves (fitting formula) for $\gamma>2$.
\label{fig:iygamma}
}\end{figure}

\begin{figure}
\vspace*{-1.0cm}
\begin{center}
   \leavevmode\psfig{figure=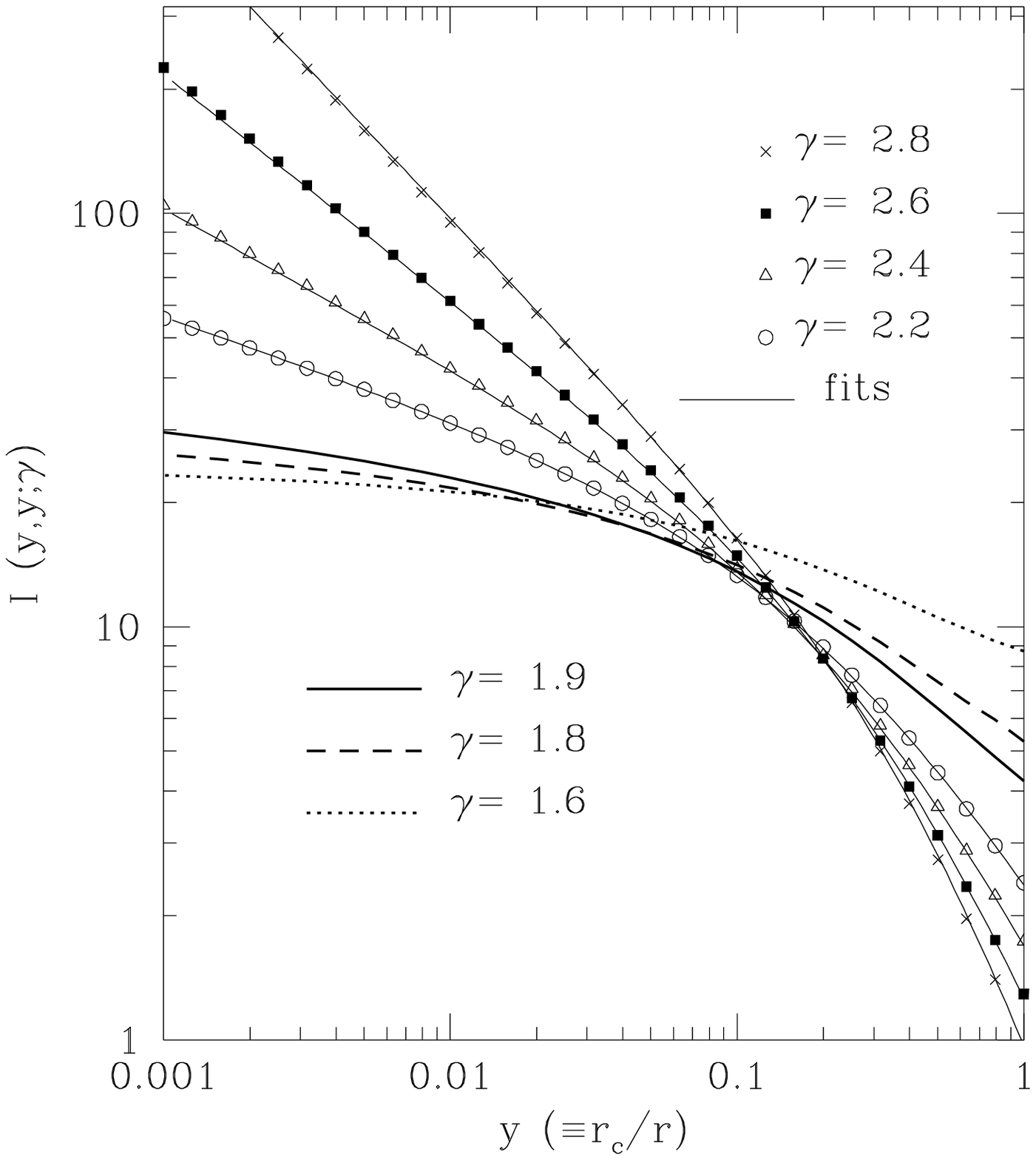,width=7.5cm}
\end{center}
\vspace*{-1.0cm}
\caption{ $I(y,y;\gamma)$ for different $\gamma$ ($r_s=r_c$).  Thick
curves indicate $\gamma<2$ while symbols (direct numerical
integration) and thin solid curves (fitting formula) for $\gamma>2$.
\label{fig:iyygamma}
}\end{figure}

If $\gamma>2$, on the other hand, the above perturbative approach does
not work and one should rely on the fully numerical integration.
Figures \ref{fig:iygamma} and \ref{fig:iyygamma} show such results;
$I(y,0;\gamma)$ and $I(y,y;\gamma)$ for various values of $\gamma$.
We find the following accurate fitting formulae for $2<\gamma<3$:
\begin{eqnarray}
I(y,0;\gamma) &=& A [ (y/y_a)^{\alpha_a} + (y/y_a)^{\beta_a} ]^{-1} , \\
I(y,y;\gamma) &=& B [ (y/y_b)^{\alpha_b} + (y/y_b)^{\beta_b} ]^{-1} ,
\end{eqnarray}
where
\begin{eqnarray}
A &=& 24.99 -36.20(\gamma-2)+13.76(\gamma-2)^2, \\
y_a &=& 0.197+0.198(\gamma-2)+0.770(\gamma-2)^2, \\
\alpha_a &=& 0.0858 +0.605(\gamma-2)+0.274(\gamma-2)^2,  \\
\beta_a &=& 0.747+0.860(\gamma-2)+0.625(\gamma-2)^2, \\
B &=& 19.23 -21.65(\gamma-2)+5.807(\gamma-2)^2, \\
y_b &=& 0.214+0.134(\gamma-2)+0.278(\gamma-2)^2, \\
\alpha_b &=& 0.116 +0.561(\gamma-2)+0.260(\gamma-2)^2,  \\
\beta_b &=& 0.946+1.003(\gamma-2)+0.469(\gamma-2)^2 .
\end{eqnarray}
At least in the range of $0.001 \simlt y\simlt 1$, the agreement is
excellent (Figs. \ref{fig:iygamma} and \ref{fig:iyygamma}) between 
the numerical results and the fitting formulae.

\section{PREDICTIONS FOR THE SMALL-SCALE VELOCITY DISPERSIONS}

\subsection{power-law models}

Let us consider the predictions for the small-scale velocity
dispersions on the basis of the CVT with the finite size effect.
With nonzero $r_s$ and $r_c$, equation (\ref{eq:cvt1d}) should be
explicitly rewritten as
\begin{eqnarray}
&& \langle v^2_{12} (r) \rangle^{1/2}_\CVT = 1460
  \, \sqrt{\Omega_0 Q_\rho \over 1.3 b_{\rm g}^2}~
   5.4^{\frac {\gamma-1.8}2} \left({r_\0
      \over 5.4\himpc} \right)^{\frac\gamma 2} \nonumber \\
  && \qquad \qquad \times
  \sqrt{I(r_c/r,r_s/r;\gamma) \over 33.2} 
\left({r \over 1\himpc }
  \right)^{\frac{2-\gamma}2} \kms .
\label{eq:cvtpowerlaw}
\end{eqnarray}
Due to the $r$-dependence of $I(r_c/r,r_s/r;\gamma)$, $\langle
v^2_{12} (r) \rangle^{1/2}_\CVT$ is not described by a power-law in
general, but is still proportional to $\sqrt{\Omega_0 Q_\rho
r_0^{\gamma}/b_{\rm g}^2}$.

\begin{figure}
\vspace*{-1.0cm}
\begin{center}
   \leavevmode\psfig{figure=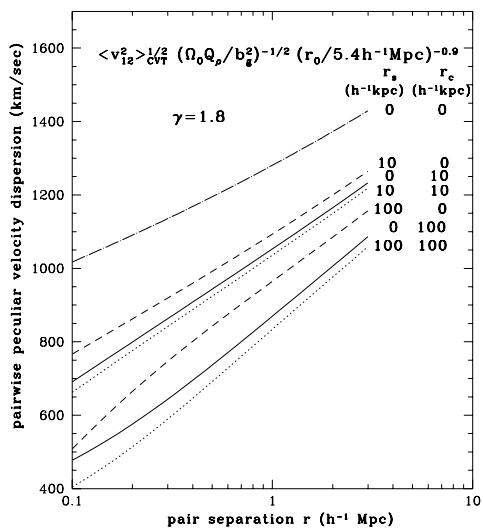,width=7.5cm}
\end{center}
\vspace*{-1.0cm}
\caption{ Relative pairwise peculiar velocity dispersions
predicted from the CVT ($\gamma=1.8$). Different curves correspond to
different sets of values for $r_s$ and $r_c$ indicated in the panel.
\label{fig:v12g18}
}\end{figure}

Figure \ref{fig:v12g18} plots $\langle v^2_{12} (r) \rangle^{1/2}_\CVT
/ \sqrt{\Omega_0 Q_\rho/b_{\rm g}^2}/(r_0/5.4\himpc)^{\gamma/2}$ for
$\gamma=1.8$.  As discussed in \S 3, even fairly small values for
$r_s$ and $r_c$ can reduce the predictions appreciably relative to the
case of the point-mass approximation ($r_s=r_c=0$). Furthermore, when
$r_c>0$, the softening of the force has negligible effect if $r_s\le
r_c$. Since the
observed two-point correlation functions of galaxies do not show any
noticeable departure from the single power-law at least for $r\simgt
0.02\himpc$, the choice of $r_s=r_c=100\hikpc$ plotted here might be
an extreme example, and already overestimate the finite size effect.

\begin{figure}
\vspace*{-1.0cm}
\begin{center}
   \leavevmode\psfig{figure=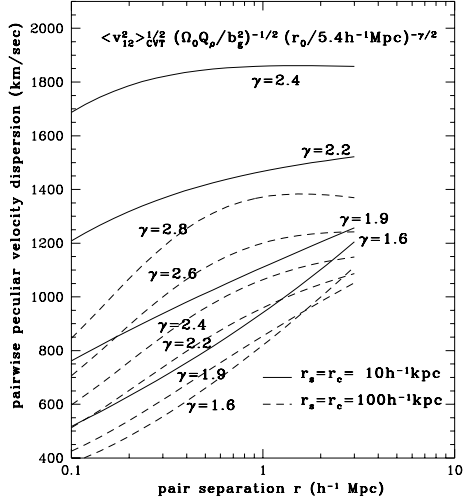,width=7.5cm}
\end{center}
\vspace*{-1.0cm}
\caption{ Relative pairwise peculiar velocity dispersions predicted
from the CVT for various $\gamma$.  Solid and dashed curves
correspond to $r_s=r_c=10\hikpc$ and $100\hikpc$, respectively.
\label{fig:v12gamma}
}\end{figure}

Similarly the results for various $\gamma$ are plotted in Figure
\ref{fig:v12gamma} for $r_s=r_c=100\hikpc$ (dashed curves) and
$r_s=r_c=10\hikpc$ (solid curves). The results for larger $\gamma$
become more sensitive to the values of $r_s$ and $r_c$. Given the fact
that the point-mass approximation fails for $\gamma >2$, this is
reasonable. It should be also pointed out here that even when
$\gamma>2$ $\langle v^2_{12} (r) \rangle^{1/2}_\CVT$ increases as $r$
because $I(r_c/r,r_s/r;\gamma)$ increases more strongly than $\propto
r^{\gamma-2}$; in the case of $r_s=r_c=100\hikpc$ and $r\simlt
1\himpc$ relevant for Figure \ref{fig:v12gamma}, $\langle v^2_{12} (r)
\rangle^{1/2}_\CVT$ is proportional to $r^{1-\gamma/2+\beta_b/2}$ =
$r^{0.48}$, $r^{0.51}$, $r^{0.56}$, and $r^{0.62}$ for $\gamma=2.2$,
$2.4$, $2.6$, and $2.8$, respectively. When $r/r_s$, $r/r_c \simgt
10$, $\langle v^2_{12} (r) \rangle^{1/2}_\CVT$ approaches $\propto
r^{1-\gamma/2+\alpha_b/2}$ and thus significantly levels off (around
$r\sim 1\himpc$ for $r_s=r_c=100\hikpc$ and $r\sim 0.1\himpc$ for
$r_s=r_c=10\hikpc$ in Fig.\ref{fig:v12cdmr}). In any case, this
implies that the scale-dependence of $\langle v^2_{12} (r)
\rangle^{1/2}_\CVT$ is fairly {\it insensitive} to the value of
$\gamma$.

\subsection{COBE normalized CDM models}

The results presented above are fairly general in the sense that they
should be valid as long as the CVT is correct and equations
(\ref{eq:hier}), (\ref{eq:cvtrs}) and (\ref{eq:xirc}) are reasonably
good approximation to the reality. Nevertheless the velocity
dispersion $\langle v^2_{12} (r) \rangle^{1/2}_\CVT$ is still
crucially dependent on the parameters including $\Omega_0$, $Q_\rho$,
$b_{\rm g}$, $r_0$, $\gamma$ as well as $r_s$ and $r_c$ which we have
discussed in details here. Therefore to proceed further one has to
assume some sort of specific models.  For this purpose cold dark
matter (CDM) models would provide good working hypotheses.  In
particular very accurate semi-analytical fitting formulae for
$\xi_\rho(r)$ have been worked out in the literature, which have been
successfully tested against high-resolution N-body simulations
(Hamilton et al. 1991; Jain, Mo \& White 1995; Peacock \& Dodds 1994,
1996; Mo, Jing \& B\"{o}rner 1996).

CDM models with the primordial scale-invariant (Harrison -Zel'dovich)
spectrum have the power spectrum $P(k)$ which is asymptotically
proportional to $k^{-3}$ for large $k$. This implies that the
two-point correlation function in these models should approach
constant on the corresponding scales; for a purely scale-free
primordial spectra of $P(k)\propto k^n$, $\xi_\rho(r)$ is proportional
to $r^{-3-n}$ in a linear regime and to $r^{-3(n+3)/(n+5)}$ in a
stable clustering (strongly nonlinear) regime.  In other words, we
naturally expect the small-scale flattening of the form
(\ref{eq:xirc}) even independently of the finite size effect of
galaxies. This is another reason why the present analysis is very
important in realistic cosmological scenarios.

Given the cosmological parameters $\Omega_0$, $\lambda_0$, and $h$,
the normalization constant of the fluctuation spectrum can be
determined by COBE data (e.g., Sugiyama 1995), and thus the
(non-linear) mass correlation function $\xi_\rho(r)$ can be
analytically calculated (Peacock \& Dodds 1996; Mo et al. 1996). We
then determine $r_0$, $\gamma$ and $r_c$ by fitting the calculated
$\xi_\rho(r)$ to equation (\ref{eq:xirc}).  So determined $r_c$ does
not represent the finite size effect of galaxies, and if it is below
$10\hikpc$, for instance, we may set $r_s=r_c=10\hikpc$ to include the
effect. Therefore as long as the hierarchical relation (\ref{eq:hier})
holds for the two- and three-point correlation functions of mass, the
only remaining parameter is $Q_\rho$. Although recent numerical and
semi-analytical studies indicate that $Q_\rho$ is weakly dependent on
the scale (Suto 1993; Matsubara \& Suto 1994; Suto \& Matsubara 1994;
Jing \& B\"{o}rner 1996; Mo et al. 1996), it is about $2\sim 3$ on
scales less than $1\himpc$.

We use the following expression for the COBE 2yr data normalization
$\sigma_{\rm COBE}(8\himpc)$ by Nakamura (1996) who performed a fit to
the numerical results by Sugiyama (1995):
\begin{eqnarray}
&& \sigma_{\rm COBE}(8\himpc) = (30\Gamma)^2 \, f(8\Gamma) 
\qquad \qquad \nonumber \\
&&\times
  \left\{ 
      \begin{array}{ll}
         \sqrt{1+1.22(1-\Omega_0)-0.266(1-\Omega_0)^2-1.23(1-\Omega_0)^3}
           & \mbox{$(\lambda_0=0)$} \\
         \Omega_0^{-0.696} & \mbox{$(\lambda_0=1-\Omega_0)$}
      \end{array}
   \right. 
 \\ \nonumber 
\end{eqnarray}
where
\begin{eqnarray}
\Gamma &\equiv& \Omega_0 h 
 \exp{\left[-\Omega_b\left(1+{\sqrt{2h}\over \Omega_0}\right)\right]}, \\
f(x) &\equiv& (18.1x^{0.2}+7.17x^{0.5}+22.1x+1.23x^2)^{-1} .
\end{eqnarray}
For definiteness we adopt the baryon density parameter
$\Omega_b=0.015h^{-2}$. Note that the COBE 2yr data give a slightly
larger normalization amplitude (roughly by 10 percent) compared with
the COBE 4yr data (Sugiyama 1996, in private communication). Our
conclusions below, however, are insensitive to this issue.

\begin{table}
\begin{center}
{Table 1: Fitted parameters for CDM models}
\begin{tabular}{rrrcccc}
\\
\tableline
$\Omega_0$ & $\lambda_0$ & $h$ & $\sigma_{\rm COBE}(8\himpc)$ &
$\gamma$ & $r_0(\himpc)$ & $r_c(\himpc)$ \\
\hline
\hline

$1$ & $0$ & $0.5$ & 1.35 & 2.002&7.67 &0.061 \\
$0.2$ & $0$ & $0.7$ & 0.27 & 1.42&0.79 &0.0  \\
$0.2$ & $0.8$ & $0.7$ & 0.78 & 1.99 &4.67 & 0.087  \\
\hline
\end{tabular} 
\end{center}
\end{table} 

\begin{figure}
\vspace*{-1.0cm}
\begin{center}
   \leavevmode\psfig{figure=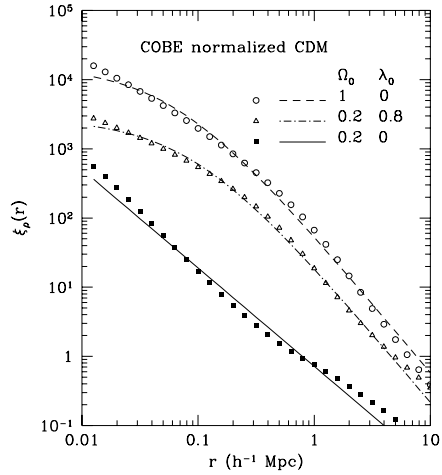,width=7.5cm}
\end{center}
\vspace*{-1.0cm}
\caption{ Two-point correlation functions of mass for the three CDM
models in Table 1 computed from analytical models (symbols) and the
power-law fits with nonzero $r_c$ (curves); $\Omega_0=1.0$ and
$\lambda_0=0.0$ (open circles and dashed curve), $\Omega_0=0.2$ and
$\lambda_0=0.0$ (filled squares and solid curve), and
$\Omega_0=0.2$ and $\lambda_0=0.8$ (open triangles and dot-dashed curve).
\label{fig:xifit}
}\end{figure}

Figure \ref{fig:xifit} plots the two-point correlation functions of
mass based on the Peacock \& Dodds (1996) model and our fits to the
form (\ref{eq:xirc}) for the CDM models in Table 1. The parameter
fitting is performed over the range $0.01\himpc \simlt r \simlt
5\himpc$.  Although the fit does not seem so good especially on small
scales, any single power-law form (i.e., $r_c=0$) yields much poorer
fits and the inclusion of the nonzero $r_c$ definitely improve the
theoretical prediction.

\begin{figure}
\vspace*{-1.0cm}
\begin{center}
   \leavevmode\psfig{figure=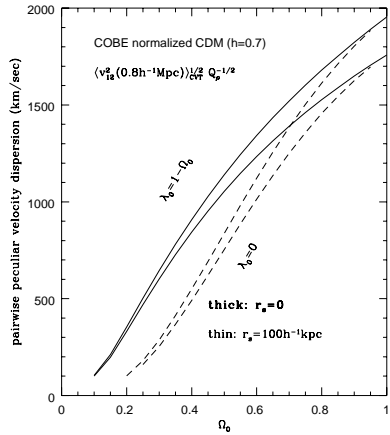,width=7.0cm}
\end{center}
\vspace*{-1.0cm}
\caption{ Relative pairwise peculiar velocity dispersions at
$r=0.8\hikpc$ as functions of $\Omega_0$ for COBE normalized CDM
models $(h=0.7)$; $\lambda_0=0$ models in dashed and
$\lambda_0=1-\Omega_0$ models in solid curves.  Thick curves indicate
for $r_s=0$, while thin curves for $r_s= 100\hikpc$.
\label{fig:v12omega}
}\end{figure}

Figure \ref{fig:v12cdmr} shows $\langle v^2_{12} (r)
\rangle^{1/2}_\CVT/\sqrt{Q_\rho}$ for the three CDM models in Table 1.
The model parameters except for $Q_\rho$ are completely determined
from theory and the fitting, and therefore the result is fairly
definite.  As mentioned above, recent numerical and semi-analytical
results suggest $Q_\rho = (2\sim 3)$ on scales less than
$1\himpc$. With this value in mind, Figure \ref{fig:v12cdmr} implies
that $\Omega_0=1$ CDM models disagree with the observed values
$(300\sim600)\kms$ at $r=1\himpc$ (Davis \& Peebles 1983; Mo, Jing \&
B\"{o}rner 1993; Fisher et al. 1994; Marzke et al. 1995 ) by a factor
of $(3\sim 6)$.  Although the conflict with the $\Omega_0=1$ model has
been pointed out frequently in the literature (e.g., Peebles 1976,
1980; Davis \& Peebles 1983; Suto 1993), our present prediction is the
most precise one from a theoretical point of view.  Finally in Figure
\ref{fig:v12omega}, we plot $\langle v^2_{12} (r=0.8\himpc)
\rangle^{1/2}_\CVT/\sqrt{Q_\rho}$ as a function of $\Omega_0$ both for
$\lambda_0=0$ and $1-\Omega_0$ models using the same procedure
described here. Clearly only the low-density models are compatible
with the observations even if one allows for a factor of $(2 \sim 3)$
uncertainty of the sample-to-sample variation (Suto 1996; Suto \& Jing
1996). It is interesting to note from the figure that softening in the
force with $r_s\approx 100 \hikpc$ has almost negligible effect on the
prediction of $\langle v_{12}^2\rangle^{1/2}$ in the three CDM models;
this is simply because these models have either $\gamma \ll 2$ or $r_c
\sim 100 \hikpc$.

\section{DISCUSSION AND CONCLUSIONS}

We have presented a realistic refinement of the cosmic virial theorem
by taking into account of the finite size effect of galaxies. The CVT
in its conventional form which treats galaxies as point particles
appreciably overestimates small-scale velocity dispersions of
galaxies.  Although this is already remarked by Peebles (1976) and in
a slightly different context by Bartlett \& Blanchard (1996), we
presented more detailed analysis on the basis of a perturbation theory
and systematic numerical calculations. In particular the former
provides a clear physical insight into the finite size effect.  Our
perturbation formulae (eqs.[\ref{eq:iy0}] -- [\ref{eq:iyy}]), which
are accurate enough [better than 20\% in accuracy for $I(y,w;\gamma)$]
for most purposes, can be easily and reliably used to correct for the
flattening of mass correlation $\xi_\rho (r)$ and the force softening
for $\gamma<2$. We have also presented accurate fitting formulae for
$I(y,w;\gamma)$ for the case of $\gamma>2$.  Moreover, we have
predicted the pairwise velocity dispersion for CDM models with the
COBE fluctuation normalization.

As our result indicates that the flattening of mass correlation
$\xi_\rho (r)$ has significant effect on the CVT, one should be very
cautious to infer $\Omega_0$ from observational measurements of
galaxy-galaxy correlation function $\xi_g(r)$ and galaxy pairwise
velocity dispersions.  Although current observations of galaxy
clustering have not revealed any noticeable flattening in $\xi_g(r)$
on small scales, non-linear biasing mechanisms may have complicated
the relation between $\xi_g(r)$ and $\xi_\rho (r)$. Moreover, the
statistical significance of galaxy clustering on very small scales
$r\simlt 200 \hikpc$ is still rather poor. In fact, intrinsic
flattening of correlation functions on small scales is expected
generally for CDM models with a Harrison -Zel'dovich primordial
spectrum. We have quantitatively pointed out the importance of taking
into account such flattening in $\xi_\rho (r)$ when one applies the
CVT, otherwise the estimated value for $\Omega_0$ would be
substantially {\it underestimated}.

\begin{figure}
\vspace*{-1.0cm}
\begin{center}
   \leavevmode\psfig{figure=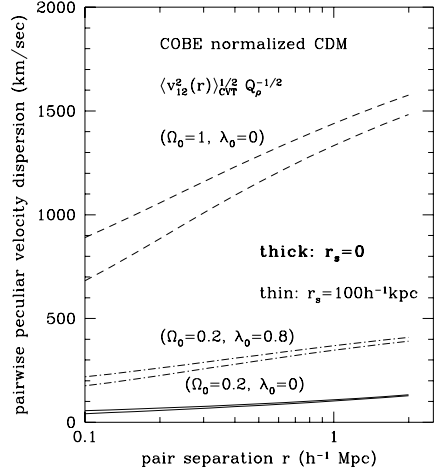,width=7.5cm} 
\end{center}
\vspace*{-1.0cm} 
\caption{ Relative pairwise peculiar velocity dispersions from the CVT
as functions of $r$ for the three CDM models in Table 1.
$\Omega_0=1.0$ and $\lambda_0=0.0$ (dashed curve), $\Omega_0=0.2$ and
$\lambda_0=0.8$ (dot-dashed curve), and $\Omega_0=0.2$ and
$\lambda_0=0.0$ (solid curve). Thick curves indicate for $r_s=0$,
while thin curves for $r_s=100\hikpc$.
\label{fig:v12cdmr} 
}\end{figure}

To avoid the complicated issue of possible non-linear biasing on small
scales, one can draw constraints on $\Omega_0$ from a comparison of
the measured pairwise velocity dispersion with our predictions for CDM
models of various $\Omega_0$ (See also Mo et al. 1996).  As long as
one relies on the COBE normalization, our CDM predictions are fairly
definite; for realistic CDM $\xi_\rho (r)$ with flattening,
fortunately, the (unknown) force softening even with $r_s \simlt 100
\hikpc$ would not make any difference on the prediction of $\langle
v_{12}^2\rangle^{1/2}$ (see \S 4.2 and Fig.\ref{fig:v12cdmr}). The
remaining two parameters which would complicate the comparison between
the predictions and observations include the normalized amplitude of
the three-point correlation functions which is unlikely to be
determined accurately from a small sample of galaxies, and a possible
velocity bias of galaxies which may result from the dynamical friction
but whose quantitative effect is still controversial in literature.

Although the present analysis provides a theoretical improvement over
the conventional CVT, the usefulness of the CVT to estimate $\Omega_0$
also relies on the quality of the observational data. In fact, Mo et
al. (1993) noted that the observed velocity dispersions vary fairly
significantly from sample to sample, and also are sensitive to the
modeling of the data. We will discuss this topic in details elsewhere
(Suto \& Jing 1996).

\bigskip

We thank Houjun Mo and Gerhard B\"orner for useful 
discussions and comments on the
manuscript, and Takahiro T. Nakamura for providing the fitting formula
for the COBE 2yr normalization.  YPJ gratefully acknowledges the
receipt of an Alexander-von-Humboldt research fellowship.  YS is
grateful to Gerhard B\"{o}rner for hospitality at Max-Planck-Institut
f\"{u}r Astrophysik, where the present work was done.  This research
was supported in part by the Grants-in-Aid by the Ministry of
Education, Science, Sports and Culture of Japan (07CE2002) to RESCEU
(Research Center for the Early Universe).

\bigskip
\bigskip

\newpage

\baselineskip=18pt
\parskip2pt
\bigskip
\centerline{\bf REFERENCES}
\bigskip

\def\apjpap#1;#2;#3;#4; {\pp#1, {#2}, {#3}, #4}
\def\apjbook#1;#2;#3;#4; {\pp#1, {#2} (#3: #4)}
\def\apjppt#1;#2; {\pp#1, #2.}
\def\apjproc#1;#2;#3;#4;#5;#6; {\pp#1, {#2} #3, (#4: #5), #6}

\apjppt Ballinger, W.E., Peacock, J.A. \& Heavens, A.F. 1996;MNRAS, in
press;
\apjpap Bartlett, J. G. \& Blanchard, A. 1996;A\&A;307;1;
\apjpap Davis, M. \& Peebles, P.J.E. 1983;ApJ;267;465;
\apjpap Fisher, K.B., Davis,M., Strauss, M.A., Yahil, A., 
  \& Huchra, J. 1994;MNRAS;267;927;
\apjpap Hamilton, A.J.S., Kumar, P., Lu, E., \& Matthews, A. 1991;
  ApJ;374;L1;
\apjpap Hamilton, A.J.S. 1992;ApJ;385;L5;
\apjpap Jain, B., Mo, H.J., \& White, S.D.M. 1995;MNRAS;276;L25;
\apjppt Jing, Y.P. \& B\"{o}rner, G. 1996;A\&A, in press;
\apjpap Kaiser, N. 1987;MNRAS;227;1;
\apjpap Marzke,R.O., Geller, M.J., da Costa, L.N., \& Huchra,J. 1995;
AJ;110;477;
\apjpap Matsubara,~T. \& Suto, Y.  1994;ApJ;420;497;
\apjppt Matsubara,~T. \& Suto, Y.  1996;ApJ October 10 issue, in press;
\apjpap Mo, H.J., Jing, Y.P., \& B\"{o}rner, G. 1993;MNRAS;264;825;
\apjppt Mo, H.J., Jing, Y.P., \& B\"{o}rner, G. 1996;MNRAS, submitted;
\apjppt Nakamura, T.T. 1996; master thesis to the University of
  Tokyo, unpublished;
\apjpap Peacock, J.A. \& Dodds, S.J. 1994;MNRAS;267;1020;
\apjpap Peacock, J.A. \& Dodds, S.J. 1996;MNRAS;280;L19;
\apjpap Peebles, P.J.E. 1976;Astrophys.Sp.Sci.;45;3;
\apjbook Peebles, P.J.E. 1980; The Large Scale Structure of the
  Universe;Princeton University Press;Princeton;
\apjpap Sugiyama,N., 1995;ApJS;100;281;
\apjpap Suto, Y.  1993;Prog.Theor.Phys.;90;1173;
\apjproc Suto, Y.  1996;in Proceedings of the inauguration conference
of Asia Pacific Center for Theoretical Physics;eds. J.W.Kim;World
Scientific;Singapore;in press;
\apjppt Suto, Y. \& Jing, Y.P. 1996;in preparation;
\apjpap Suto, Y. \& Matsubara,~T.  1994;ApJ;420;504;
\apjppt Suto, Y. \& Matsubara,~T.  1996;ApJ, submitted;

\bigskip

\end{document}